\documentclass[preprint,prb,aps,amsmath,amssymb,showpacs]{revtex4-1}
\usepackage[english]{babel}

\usepackage{graphicx}

\begin{document}

\title{Electron-phonon-scattering dynamics in ferromagnetic metals %
and its influence on ultrafast demagnetization processes}

\date{\today}

\author{Sven Essert}
\author{Hans Christian Schneider}
\email{hcsch@physik.uni-kl.de}
\affiliation{Department of Physics and Research Center OPTIMAS, %
University of Kaiserslautern, P.O. Box 3094, 67653 Kaiserslautern, Germany}
 
\begin{abstract}
We theoretically investigate spin-dependent carrier dynamics due to
the electron-phonon interaction after ultrafast optical excitation in
ferromagnetic metals.  We calculate the electron-phonon matrix
elements including the spin-orbit interaction in the electronic wave
functions and the interaction potential.  Using the matrix elements in
Boltzmann scattering integrals, the momentum-resolved carrier
distributions are obtained by solving their equation of motion
numerically. We find that the optical excitation with realistic laser
intensities alone leads to a negligible magnetization change, and that
the demagnetization due to electron-phonon interaction is mostly due
to hole scattering. Importantly, the calculated demagnetization
quenching due to this Elliot-Yafet type depolarization mechanism is
not large enough to explain the experimentally observed result. We
argue that the ultrafast demagnetization of ferromagnets does not
occur exclusively via an Elliott-Yafet type process, i.e., scattering
in the presence of the spin-orbit interaction, but is influenced to a
large degree by a dynamical change of the band structure, i.e., the
exchange splitting.
\end{abstract}
\maketitle

\section{Introduction}
It was first demonstrated more than ten years ago that the
magnetization of ferromagnets can be ``quenched'' on ultrashort time
scales after ultrafast optical excitation.~\cite{Beaurepaire1996} Apart from the
possibilities for the ultrafast manipulation of ferromagnetism in
applications, this observation raised the question how demagnetization dynamics in ferromagnets
on a time scale of a few hundred femtoseconds can be understood.

Aside from the phenomenological three-temperature model,~\cite{Beaurepaire1996,Koopmans:2010ds} which leads to quite
successful comparison with experiment, there are
several theoretical models and experimental results that try to explain aspects of the underlying microscopic dynamics. For instance,  the analysis of  X-ray
magnetic circular dichroism measurements suggested
that the orbital magnetic moment does not play a prominent role in
the demagnetization dynamics.~\cite{Stamm2007,Carva2009} The authors of Ref.~\onlinecite{Stamm2007} concluded that an ultrafast
spin-lattice coupling should be operative to explain the results. It
has also been argued based on experimental results~\cite{Carpene2008} that the excitation
of magnons should play an important role. On the theory side, magnetic switching due to electronic transitions during the duration of a pump laser pulse has been analyzed in ferromagnets~\cite{ZhangHuebner:2000prl, Zhang2009} as well as in oxides (including phonons),~\cite{Lefkidis:2009jmmm}
and the
Landau-Lifshitz-Bloch equations have been used to describe the magnetic
dynamics.~\cite{Atxitia:2010bv}

Perhaps the most popular microscopic explanations of the effect
involve variations of the so-called Elliott-Yafet mechanism, in which
demagnetization (or depolarization in semiconductors) is due to
incoherent scattering of carriers between states that are spin-mixed
due to the spin-orbit interaction. Electron-electron scattering is a
possible candidate as the underlying scattering
mechanism,~\cite{Krauss:2009gc} but the focus is usually on the
effects of electron-phonon scattering in
quasi-equilibrium.~\cite{Koopmans:2005dz,Walowski:2008dg,Steiauf:2009ca,Steiauf:2010ea}
In addition, superdiffusive transport processes can contribute to the
measured Kerr-effect signal because minority and majority electrons
may simply leave the probe area at different
speeds.~\cite{Battiato:2010br}

In our opinion, it has so far been impossible rule out a single one of these mechanisms, let alone to pinpoint the dominant one. As a first step in this direction, we analyze a parameter-free
microscopic model for ultrafast demagnetization and compare it with
experiment. To keep things simple while allowing a conclusive
statement, we exclusively treat the effect of the optical excitation
and electron-phonon scattering at the level of Boltzmann scattering
integrals while neglecting dynamical changes in the band structure,
i.e., the exchange splitting, in the course of the dynamics.  We
evaluate the model using ab-initio results for the simple ferromagnets nickel and
iron. Comparison of the results of the present paper with experimental data, which are available from many
different measurements, will show that a model without band structure
changes yields a demagnetization that is too small.

This paper is organized as follows. In Sec.~II we present the
dynamical equations for the carrier distribution functions and show
how we calculate the electron-phonon and dipole matrix elements using
a first-principles approach. In Sec.~III we discuss the numerical
results of this model and show that the demagnetization using
realistic parameters for the ultrashort-pulse excitation is due to
hole dynamics, but too small to agree with experiment. A qualitative
consideration shows that this conclusion should not be altered by
including additional scattering processes. Sec.~IV contains the
conclusions and the Appendix describes details of our numerical evaluation of
the dynamical equations.

\section{Theory}

\subsection{Dynamical equations}

The basic idea of this paper is to integrate the dynamical equations
of motions for the band- and momentum-resolved carrier distributions
$n_{\vec{k}}^{\mu}(t)$.  Our model
includes the incoherent scattering dynamics due to the electron-phonon
interaction, as well as the optical excitation, so that the general
form of the dynamical equation is
\begin{equation}
\frac{\partial}{\partial t} n_{\vec{k}}^{\mu} = 
\frac{\partial}{\partial t} n_{\vec{k}}^{\mu}\Bigr|_{\text{opt}}+
\frac{\partial}{\partial t} n_{\vec{k}}^{\mu}\Bigr|_{\text{e-p}}.
\label{eq:dndt}
\end{equation}
The optical excitation is given by
\begin{equation}
\frac{\partial}{\partial t} n_{\vec{k}}^{\mu}\Bigr|_{\text{opt}} =
\frac{2\pi}{\hbar}\sum_{\nu (\neq\mu)} 
    \bigl|\vec{d}_{\vec{k}}^{\mu\nu}\cdot\vec{E}(t)\bigr|^{2}
    g(\epsilon_{\vec{k}}^{\mu}-\epsilon_{\vec{k}}^{\nu}-\hbar\omega_{L})
    \bigl[n_{\vec{k}}^{\nu}-n_{\vec{k}}^{\mu}\bigr].
\end{equation}
Here, $\epsilon_{\vec{k}}^{\mu}$ is the energy of a carrier in a
single-particle state $\psi_{\vec{k}}^{\mu}$ with band index
$\mu$ and momentum $\vec{k}$. The dipole matrix element for a
transition connecting two such states is denoted by
$\vec{d}_{\vec{k}}^{\mu\nu}=\langle\psi_{\vec{k}}^{\mu}|\,
e\vec{r}\,|\psi_{\vec{k}}^{\nu}\rangle$. The optical excitation is
characterized by the dynamical electric field amplitude $\vec{E}(t)$, a central laser  frequency $\omega_{\mathrm{L}}$,
and the function~$g(\epsilon)$ that includes the spectral profile of
the laser pulse.

The electron-phonon contribution to the carrier dynamics at the level of
Boltzmann scattering integrals reads
\begin{widetext}
\begin{equation}
\frac{\partial}{\partial t} n_{\vec{k}}^{\mu}\Bigr|_{\text{e-p}}=
   \frac{2\pi}{\hbar}\sum_{\vec{k}',\nu}\Bigl[ 
   w(\vec{k}',\nu\rightarrow\vec{k},\mu)
       n_{\vec{k}'}^{\nu}\bigl(1-n_{\vec{k}}^{\mu}\bigr) \\
  - w(\vec{k},\mu\rightarrow\vec{k}',\nu)n_{\vec{k}}^{\mu}
       \bigl(1-n_{\vec{k}'}^{\nu}\bigr)\Bigr],
\end{equation}
where the scattering rates $w(\vec{k}',\nu\rightarrow\vec{k},\mu)$
from state $\psi_{\vec{k}'}^{\nu}$ to $\psi_{\vec{k}}^{\mu}$ 
\begin{equation}
\begin{split}
w(\vec{k}',\nu\rightarrow\vec{k},\mu)=
\sum_{\lambda}\Bigl[&
  \bigl|M_{+}^{\vec{k}-\vec{k}',\lambda}(\vec{k}',\nu\rightarrow\vec{k},\mu)\bigr|^{2} 
  \tilde{n}_{\vec{k}-\vec{k}'}^{\lambda} \,
      \delta(\epsilon_{\vec{k}'}^{\nu}-\epsilon_{\vec{k}}^{\mu}+\hbar\omega_{\vec{k}-\vec{k}'}^{\lambda}) \\
 +&\bigl|M_{-}^{\vec{k}'-\vec{k},\lambda}(\vec{k}',\nu\rightarrow\vec{k},\mu)\bigr|^{2}
    (\tilde{n}_{\vec{k}'-\vec{k}}^{\lambda}+1)\,
    \delta(\epsilon_{\vec{k}'}^{\nu}-\epsilon_{\vec{k}}^{\mu}-\hbar\omega_{\vec{k}'-\vec{k}}^{\lambda})\Bigr]
\end{split}
\label{eq:def_w}
\end{equation}
have contributions from absorption (``$+$'') and emission (``$-$'')
processes. The (angular) frequency of a phonon mode $\lambda$ is
designated by $\omega_{\vec{q}}^{\lambda}$ and its occupation at
quasi-momentum $\vec{q}$ by $\tilde{n}_{\vec{q}}^{\lambda}$. The
electron-phonon interaction matrix elements
$M_{{\pm}}^{\vec{q},\lambda}(\vec{k}',\nu\rightarrow\vec{k},\mu)$
result from the change of the electron-lattice interaction energy due
to the vibrational motion of the nuclei. For small displacements they
are given by
\begin{equation}
M_{{\pm}}^{\vec{q},\lambda}(\vec{k}',\nu\rightarrow\vec{k},\mu)
= \sqrt{\frac{\hbar}{2MN\omega_{\vec{q}}^{\lambda}}}
\sum_{j}e^{\pm i\vec{q}\cdot\vec{R}_{j}}
\Bigl\langle\psi_{\vec{k}}^{\mu}\Bigr|\vec{\pi}_{\vec{q}}^{\lambda}\cdot\frac{\partial}{\partial \vec{R}_{j}}
V(\vec{r};\{\vec{R}_{i}\})\Bigl|\psi_{\vec{k}'}^{\nu}\Bigr\rangle. 
\label{eq:elph_me}
\end{equation}
\end{widetext}
The electron-lattice interaction potential
$V(\vec{r};\{\vec{R}_{i}\})$ depends on the electron position
$\vec{r}$ and, in principle, on the set of the positions of the
nuclei $\{\vec{R}_{i}\}$ in the crystal composed of $N$ unit cells
with atomic mass~$M$. The polarization vector of the phonon mode
$(\vec{q},\lambda)$ is denoted by $\vec{\pi}_{\vec{q}}^{\lambda}$. The
upper and lower signs in the exponential are associated with the
phonon absorption
$M_{+}^{\vec{q},\lambda}(\vec{k}',\nu\rightarrow\vec{k},\mu)$ and
emission $M_{-}^{\vec{q},\lambda}(\vec{k}',\nu\rightarrow\vec{k},\mu)$
matrix elements, respectively.

In our calculation, we assume that the phonon occupation numbers
are time independent and remain at their equilibrium values 
\begin{equation}
\tilde{n}_{\vec{q}}^{\lambda}=\frac{1}{e^{\hbar\omega_{\vec{q}}^{\lambda}/k_{B}T_{0}}-1}.
\end{equation}
This amounts to a bath assumption for the phonon system, and in this
paper we fix its temperature at $T_{0} = 300$\,K, i.e., the temperature
of the unexcited system in most studies of demagnetization dynamics.

From the dynamical electronic occupation
numbers calculated according to Eq.~(\ref{eq:dndt}), we obtain the time-dependent magnetization
by
\begin{equation}
M(t)=\frac{2\mu_{B}}{\hbar}\sum_{\vec{k}}\sum_{\mu}
   \left\langle S_{z}\right\rangle _{\vec{k}}^{\mu}n_{\vec{k}}^{\mu}(t),
\end{equation}
with the single-particle spin expectation value 
\begin{equation}
\left\langle S_{z}\right\rangle _{\vec{k}}^{\mu}
=\bigl\langle\psi_{\vec{k}}^{\mu}\bigr|\hat{S}_{z}\bigl|\psi_{\vec{k}}^{\mu}\bigr\rangle
\end{equation}
in the Bloch state $\bigl|\psi_{\vec{k}}^{\mu}\bigr\rangle$ and the
Bohr magneton $\mu_{B}$. In writing these relations, we have chosen
the $z$-direction as direction of the ferromagnetic
polarization. Orbital angular-momentum contributions to the
magnetization are neglected as their influence on the magnetization of
elementary ferromagnets is small.

\subsection{Electron-phonon matrix elements}

The numerical evaluation of Eq.~\eqref{eq:dndt} requires as input
material properties, in particular, the electronic band structure
$\epsilon_{\vec{k}}^{\mu}$, the spin expectation value of the
single-particle states $\langle S_{z}\rangle_{\vec{k}}^{\mu}$, the
dipole transition matrix elements $\vec{d}_{\vec{k}}^{\mu\nu}$, the
phonon dispersion $\omega_{\vec{q}}^{\lambda}$, and, importantly, the
electron-phonon matrix elements
$M_{{\pm}}^{\vec{q},\lambda}(\vec{k}',\nu\rightarrow\vec{k},\mu)$.  We
obtain these quantities from density-functional theory to avoid the
introduction of adjustable parameters. To this end, we employ the
augmented spherical wave (ASW) method~\cite{Williams1979} as
described in the monograph Ref.~\onlinecite{Kubler2000} (see also
Ref.~\onlinecite{Eyert2007a}). The implementation of the ASW method used by us was developed in
the K\"ubler group and relies on the scalar relativistic and
local spin-density approximations. It includes spin-orbit
coupling in a second variational correction.

The starting point for the calculation of the matrix elements is
the representation of the wave function of a single-particle state
with band index $\mu$ and momentum $\vec{k}$ in the ASW basis. Since
we employ the \emph{atomic sphere approximation} (ASA) it is usually
sufficient to know the wave functions inside the atomic spheres where
they are given by
\begin{equation}
\psi_{\vec{k}}^{\mu}(\vec{r})
=\sum_{L,\sigma}\bigl[C_{L\sigma}^{\mu}(\vec{k})i^{l}\tilde{h}_{l\sigma}(r)
                    +A_{L\sigma}^{\mu}(\vec{k})i^{l}\tilde{j}_{l\sigma}(r)\bigr]
     Y_{L}(\hat{r})\chi_{\sigma}
\end{equation}
where $\tilde{h}_{l\sigma}(r)$ and $\tilde{j}_{l\sigma}(r)$ are
\emph{augmented spherical waves}, $Y_{L}(\hat{r})$
spherical harmonics, and $\chi_{\sigma}$ Pauli
spinors. Here, $L=(l,m)$ is a multi-index that includes both the angular
momentum and the magnetic quantum number.
The relatively simple expression given here is only
valid for materials with basis consisting of a single atom, such as the simple
ferromagnets investigated in the present paper. The spherical wave functions,
together with the coefficients $A_{L\sigma}^{\mu}(\vec{k})$ and
$C_{L\sigma}^{\mu}(\vec{k})$, are calculated self-consistently during
the iterative solution of the Kohn-Sham equations. 

For the evaluation of
Eq.~\eqref{eq:elph_me}, we employ the so-called \emph{rigid ion
  approximation}. That is, we assume that we can write the
lattice-configuration dependence of the electron-phonon interaction potential 
\begin{equation}
V(\vec{r};\{\vec{R}_{i}\})=\sum_{i}v(\vec{r}-\vec{R}_{i})
\end{equation}
as a superposition of the on-site potentials $v(\vec{r})$, see below, Eq.~\eqref{eq:v_withso}. We also assume
that the potential $v$ vanishes outside the atomic sphere. The rigid ion
approximation is known to give a quite realistic description of the
electron-phonon coupling in transition metals.~\cite{Grimvall1981}

Equation~\eqref{eq:elph_me} can then be simplified to yield
\begin{equation}
M_{{\pm}}^{\vec{q},\lambda}(\vec{k}',\nu\rightarrow\vec{k},\mu)
=-\sqrt{\frac{\hbar}{2MN\omega_{\vec{q}}^{\lambda}}}\delta_{\vec{k}'\pm\vec{q}-\vec{k},\vec{G}} 
 \int_{\text{UC}}d^{3}r\,\psi_{\vec{k}}^{\mu*}(\vec{r})\bigl[\vec{\pi}_{\vec{q}}^{\lambda}\cdot
\nabla v(\vec{r}) \bigr] \psi_{\vec{k}'}^{\nu}(\vec{r}),
\label{eq:melph_simpl}
\end{equation}
where UC denotes an integral over the unit cell, which due to the ASA is assumed to be spherical.
For the on-site potential experienced by the electrons, we include the
spin-averaged radial Kohn-Sham potential $V_{\text{eff}}(r)$ as well as the spin-orbit interaction
\begin{equation}
v(\vec{r})=V_{\text{eff}}(r)
+\frac{\hbar}{\left(2mc\right)^{2}}\frac{1}{r}\frac{dV_{\text{eff}}(r)}{dr}
\vec{\sigma}\cdot\left(\vec{r}\times\vec{p}\right). 
\label{eq:v_withso}
\end{equation}

The additional spin-orbit term is often neglected for the
electron-phonon interaction, even though it has been shown to be of importance for
spin-relaxation in materials with time-inversion
symmetry.~\cite{Yafet1963} Not much is known about the influence of
this term in ferromagnets where the time-inversion symmetry is
broken. We therefore calculate the matrix element 
with and without the spin-orbit contribution and show that
our final results on ultrafast demagnetization are not
qualitatively influenced by the inclusion of the second term.

We first give the result for the calculation \emph{without} the spin-orbit
contribution in Eq.~\eqref{eq:v_withso}. In this case we can directly
evaluate the integral over the unit cell in
Eq.~\eqref{eq:melph_simpl}:
\begin{widetext}
\begin{align}
\label{melph-numerical}
\int_{\text{UC}}d^{3}r\,\psi_{\vec{k}}^{\mu*}(\vec{r})
    \nabla V_{\text{eff}}(r)\psi_{\vec{k}'}^{\nu}(\vec{r})\nonumber
=&\int_{\text{UC}}d^{3}r\,\psi_{\vec{k}}^{\mu*}(\vec{r})
     \frac{dV_{\text{eff}}(r)}{dr}\hat{r}\psi_{\vec{k}'}^{\nu}(\vec{r})\\
=\sum_{\sigma}\sum_{L,L'}\vec{G}_{LL'}
\Biggl[&A_{L\sigma}^{\mu*}(\vec{k})A_{L'\sigma}^{\nu}(\vec{k}')\langle\tilde{j}|
    \frac{dV_{\text{eff}}}{dr}|\tilde{j}\rangle_{l\sigma,l'\sigma} 
+C_{L\sigma}^{\mu*}(\vec{k})A_{L'\sigma}^{\nu}(\vec{k}')\langle\tilde{h}|
     \frac{dV_{\text{eff}}}{dr}|\tilde{j}\rangle_{l\sigma,l'\sigma} \nonumber \\
+&A_{L\sigma}^{\mu*}(\vec{k})C_{L'\sigma}^{\nu}(\vec{k}')\langle\tilde{j}|
     \frac{dV_{\text{eff}}}{dr}|\tilde{h}\rangle_{l\sigma,l'\sigma} 
+C_{L\sigma}^{\mu*}(\vec{k})C_{L'\sigma}^{\nu}(\vec{k}')\langle\tilde{h}|
     \frac{dV_{\text{eff}}}{dr}|\tilde{h}\rangle_{l\sigma,l'\sigma}\Biggr].
\end{align}
\end{widetext}
The radial matrix elements
\begin{equation}
\langle\tilde{f}|\frac{dV_{\text{eff}}}{dr}|\tilde{g}\rangle_{l\sigma,l'\sigma'}
 =\left(-1\right)^{l}i^{l+l'}\int_{0}^{r_{K}}r^{2}\tilde{f}_{l\sigma}(r)\frac{dV_{\text{eff}}(r)}{dr}\tilde{g}_{l'\sigma'}(r)dr 
\label{eq:f_dVdr}
\end{equation}
can be calculated by integrating the gradient of the Kohn-Sham potential and 
\begin{equation}
\vec{G}_{LL'}=\int d\Omega\, Y_{L}^{*}(\hat{r})\hat{r}Y_{L'}(\hat{r})\label{eq:G}
\end{equation}
can be evaluated in terms of Gaunt coefficients. \cite{Eyert2007a}

The calculation of the electron-phonon interaction matrix element~\eqref{eq:melph_simpl}  \emph{including the spin-orbit contribution} could, in principle, be achieved by evaluating the integral
\begin{equation}
\label{melph-onlyso}
\int_{\text{UC}}d^{3}r\,\psi_{\vec{k}}^{\mu*}(\vec{r})
\nabla\left[\frac{\hbar}{\left(2mc\right)^{2}}
\frac{1}{r}\frac{dV_{\text{eff}}(r)}{dr}
\vec{\sigma}\cdot\left(\vec{r}\times\vec{p}\right)\right]
\psi_{\vec{k}'}^{\nu}(\vec{r})
\end{equation}
and adding it to Eq.~\eqref{melph-numerical}. However,  straightforward numerical evaluation of Eq.~\eqref{melph-onlyso} runs into difficulties
because of the strong divergence of the integrand for
$r\rightarrow 0$. We circumvent this problem by calculating the complete matrix element~\eqref{eq:melph_simpl} by rewriting the
gradient of the potential including the spin-orbit interaction [Eq.~\eqref{eq:v_withso}] as a commutator
\begin{equation}
\label{gradv-commutator}
\nabla v(\vec{r})=\frac{i}{\hbar}\left[\vec{p},H_{\text{eff}}\right]
\end{equation}
with the Hamiltonian $H_{\text{eff}}=\frac{p^2}{2m}+v(\vec{r})$. For
the evaluation of matrix elements of $H_\text{eff}$, we will assume
that it produces the energy eigenvalues $\epsilon_{\vec{k}}^{\mu}$
when acting on the corresponding Kohn-Sham eigenvector, even though
the eigenenergies and eigenvectors are computed using scalar
relativistic corrections to $H_\text{eff}$. If we now assume
completeness of the ASW basis, Eq.~\eqref{gradv-commutator} may be used for a reformulation in
terms of the momentum matrix elements
\begin{widetext}
\begin{equation}
\begin{split}
\int_{\text{UC}}d^{3}r\,\psi_{\vec{k}}^{\mu*}(\vec{r})
   \nabla v(\vec{r})\psi_{\vec{k}'}^{\nu}(\vec{r}) 
&=\frac{i}{\hbar}\int_{\text{UC}}d^{3}r\,\psi_{\vec{k}}^{\mu*}(\vec{r})
   \left[\vec{p}\,H_{\text{eff}}-H_{\text{eff}}\vec{p}\right]\psi_{\vec{k}'}^{\nu}(\vec{r}) \\
&=\frac{i}{\hbar}\epsilon_{\vec{k}'}^{\nu}\bigl\langle\psi_{\vec{k}}^{\mu}
  \big|\vec{p}\big|\psi_{\vec{k}'}^{\nu}\bigr\rangle_{\text{UC}}
-\frac{i}{\hbar}\sum_{\eta}\epsilon_{\vec{k}'}^{\eta}
  \bigl\langle\psi_{\vec{k}}^{\mu}\big|\psi_{\vec{k}'}^{\eta}\bigr\rangle_{\text{UC}}
  \bigl\langle\psi_{\vec{k}'}^{\eta}\big|\vec{p}\big|\psi_{\vec{k}'}^{\nu}\bigr\rangle_{\text{UC}},
\end{split}
\end{equation}
where the momentum matrix elements
$\bigl\langle\psi_{\vec{k}}^{\mu}\big|\vec{p}\big|\psi_{\vec{k}'}^{\nu}\bigr\rangle_{\text{UC}}$
are calculated in the ASA using  a consistent method developed by Oppeneer et al.~\cite{Oppeneer1992}  Since
$\nabla v(\vec{r})$ is a hermitian operator on
the unit cell, we can derive the following expression that is more symmetric with
regard to initial and final states:
\begin{equation}
\begin{split}
\int_{\text{UC}}d^{3}r&\psi_{\vec{k}}^{\mu*}(\vec{r})
     \nabla v(\vec{r})\psi_{\vec{k}'}^{\nu}(\vec{r}) \\
&=\frac{i}{2\hbar}\Biggl[\int_{\text{UC}}d^{3}r\,\psi_{\vec{k}}^{\mu*}(\vec{r})
    \left[\vec{p}\,H_{\text{eff}}-H_{\text{eff}}\vec{p}\right]\psi_{\vec{k}'}^{\nu}(\vec{r})
        +\left(\int_{\text{UC}}d^{3}r\,\psi_{\vec{k}'}^{\nu*}(\vec{r})
    \left[\vec{p}\,H_{\text{eff}}-H_{\text{eff}}\vec{p}\right]
       \psi_{\vec{k}}^{\mu}(\vec{r})\right)^{*}\Biggr]\\
&=\frac{i}{2\hbar}\Biggl[\epsilon_{\vec{k}'}^{\nu}
       \bigl\langle\psi_{\vec{k}}^{\mu}\big|\vec{p}\big|\psi_{\vec{k}'}^{\nu}\bigr\rangle_{\text{UC}}
        +\epsilon_{\vec{k}}^{\mu}
        \bigl\langle\psi_{\vec{k}'}^{\nu}\big|\vec{p}\big|\psi_{\vec{k}}^{\mu}\bigr\rangle_{\text{UC}}^{*}
         -\sum_{\eta}\epsilon_{\vec{k}'}^{\eta}
          \bigl\langle\psi_{\vec{k}}^{\mu}\big|\psi_{\vec{k}'}^{\eta}\bigr\rangle_{\text{UC}}
           \bigl\langle\psi_{\vec{k}'}^{\eta}\big|\vec{p}\big|\psi_{\vec{k}'}^{\nu}\bigr\rangle_{\text{UC}}\\
&\qquad\mbox{}-\sum_{\eta}\epsilon_{\vec{k}}^{\eta}
       \bigl\langle\psi_{\vec{k}'}^{\nu}\big|\psi_{\vec{k}}^{\eta}\bigr\rangle_{\text{UC}}^{*}
    \bigl\langle\psi_{\vec{k}}^{\eta}\big|\vec{p}\big|\psi_{\vec{k}}^{\mu}\bigr\rangle_{\text{UC }}^{*}\Biggr]
\end{split}
\end{equation}
\end{widetext}
Although our assumption of completeness may yield the matrix element
only with a certain error because the ASW method uses a rather small
number of basis functions, the qualitative conclusions discussed in
the next section are not affected by this. 

In the numerical
calculations, we typically used a $\vec{k}$-point grid of about $2000$
points in the irreducible wedge of the band structure, and the dynamical
equations were solved on the same grid (see the Appendix~\ref{sec:nummeth}
for details on the numerical method). Experimental values for the
lattice constants \cite{Gray1972} were used. The phonon dispersion was calculated with \textsc{Quantum Espresso}~\cite{Giannozzi2009} in the same way as by Dal Corso et
al.~\cite{Corso2000} The latter paper shows that phonon dispersions
obtained in this approach are in good agreement with experimental data.

\subsection{Dipole matrix-elements} %

The dipole matrix elements are calculated by reformulating them in terms of the momentum matrix elements
\begin{equation}
\vec{d}_{\vec{k}}^{\mu\nu}=\bigl\langle\psi_{\vec{k}}^{\mu}\big|e\vec{r}\big|\psi_{\vec{k}}^{\nu}\bigr\rangle
=\frac{ie\hbar}{m\left(\epsilon_{\vec{k}}^{\nu}-\epsilon_{\vec{k}}^{\mu}\right)}\Bigl\langle\psi_{\vec{k}}^{\mu}\Big|\vec{p}+\frac{\hbar}{4mc^{2}}\left(\vec{\sigma}\times\nabla V_{\text{eff}}(r)\right)\Big|\psi_{\vec{k}}^{\nu}\Bigr\rangle,
\end{equation}
where the momentum matrix elements
$\bigl\langle\psi_{\vec{k}}^{\mu}\big|\vec{p}\big|\psi_{\vec{k}}^{\nu}\bigr\rangle$
are again calculated according to Oppeneer et al. \cite{Oppeneer1992} The contribution of the spin-orbit interaction is usually
neglected. We include it here as it may directly contribute to
spin flips, even though our numerical results in the end will show that the difference
is insignificant. It can be calculated from the
wavefunctions as follows:
\begin{widetext}
\begin{equation}
\begin{split}
&\Bigl\langle\psi_{\vec{k}}^{\mu}\Big|\left(\vec{\sigma}\times  \nabla V_{\text{eff}}(r)\right)\Big|\psi_{\vec{k}}^{\nu}\Bigr\rangle 
=\int_{\text{UC}}d^{3}r\,\psi_{\vec{k}}^{\mu*}(\vec{r})\left(\vec{\sigma}\times\hat{r}\right)\frac{dV_{\text{eff}}(r)}{dr}\psi_{\vec{k}'}^{\nu}(\vec{r}) \\
&=\sum_{\sigma,\sigma'}\!\sum_{L,L'}\left[\left(\chi_{\sigma}^{T}\vec{\sigma}\chi_{\sigma'}\right)\times\vec{G}_{LL'}\right] 
\Biggl[A_{L\sigma}^{\mu*}(\vec{k})A_{L'\sigma'}^{\nu}(\vec{k}')\langle\tilde{j}|\frac{dV_{\text{eff}}}{dr}|\tilde{j}\rangle_{l\sigma,l'\sigma'} 
+C_{L\sigma}^{\mu*}(\vec{k})A_{L'\sigma'}^{\nu}(\vec{k}')\langle\tilde{h}|\frac{dV_{\text{eff}}}{dr}|\tilde{j}\rangle_{l\sigma,l'\sigma'} \\
&\qquad+A_{L\sigma}^{\mu*}(\vec{k})C_{L'\sigma'}^{\nu}(\vec{k}')\langle\tilde{j}|\frac{dV_{\text{eff}}}{dr}|\tilde{h}\rangle_{l\sigma,l'\sigma'} 
+C_{L\sigma}^{\mu*}(\vec{k})C_{L'\sigma'}^{\nu}(\vec{k}')\langle\tilde{h}|\frac{dV_{\text{eff}}}{dr}|\tilde{h}\rangle_{l\sigma,l'\sigma'}\Biggr]\ ,
\end{split}
\end{equation}
\end{widetext}
cf.~Eq.~\eqref{eq:f_dVdr} for the definition of the matrix elements involving $V_{\mathrm{eff}}$.
\section{Results}

In this section, we present numerical results obtained from the
solution of the dynamical equation~\eqref{eq:dndt} and some
qualitative considerations on the role of scattering processes in 
ultrafast demagnetization dynamics. For the calculations we
use matrix elements computed as described in the
previous sections. Further details of our numerical implementation of the dynamics are
included in the Appendix.

\subsection{Optical excitation}\label{sec:opt_exc}

\begin{figure}
\includegraphics[width=0.6\columnwidth]{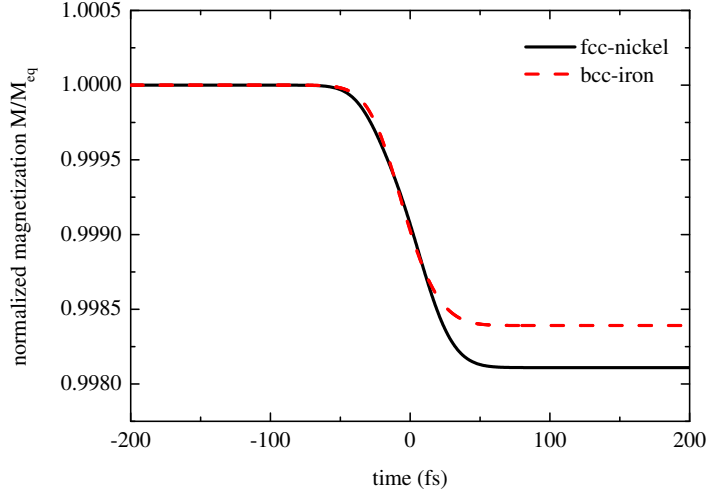}
\caption{\label{fig:simonlyopt}(Color online) Magnetization dynamics
  due to optical excitation alone. The magnetization is normalized to its
  equilibrium value.}
\end{figure}

We first examine the excitation process by the ultrashort optical
pulse without including the electron-phonon interaction. We model a
homogeneous excitation of a ferromagnetic metal by a laser pulse with gaussian temporal shape, full width at half
maximum of 50\,fs, and a spectral width of 100\,meV at a central photon
energy of $\hbar\omega=1.55$\,eV. These parameters, as well as the pulse intensity of
4\,mJ/cm$^2$, are chosen to match typical
experimental excitation conditions. To determine the electric field amplitude present in the material we note that the chosen intensity corresponds to an electric field
amplitude of $E_0=7.5\times10^8$\,V/m in vacuum. Reflection at the
surface as well as the optical density of the material lead to a
reduction of the field amplitude in the material
\begin{equation}
E_0'=E_0\frac{2}{\sqrt{(1+n)^2+\kappa^2}},
\end{equation}
where $n$ and $\kappa$ denote the real and the imaginary part of the
refractive index, respectively. Taking $n = 2.22$, $\kappa = 4.90$ for nickel
and $n=2.92 $, $\kappa = 3.36$ for iron, this leaves us with
an amplitude in the material $E_0'$ of $2.6\times10^8$\,V/m for nickel
and $2.9\times10^8$\,V/m for iron. \cite{Johnson1974}
We take these values of the field amplitude as constant throughout the sample for the calculation and neglect the attenuation due to absorption in the material as well as additional reflection/absorption due to oxide and protection layers. We therefore overestimate the field amplitude present in samples used for the experimental determination of the magnetization dynamics.

The optical excitation contribution alone, i.~e., the second term in
the dynamical equation~(\ref{eq:dndt}), leads to the magnetization dynamics shown in
Fig.~\ref{fig:simonlyopt}. This result should be compared to
experimental values for a pulse energy density of
4\,mJ/cm$^2$, such as those reported in Refs.~\onlinecite{Krauss:2009gc} and~\onlinecite{Carpene2008} where a ``quenching'' of the magnetization down to
40\,\% and 80\,\% of the equilibrium magnetization was found for nickel
and iron, respectively. It is clear from Fig.~\ref{fig:simonlyopt} that the magnetization change computed including only the incoherent optical excitation at the photon energy of 1.55\,eV is orders of magnitude smaller than the one observed in experiment. 

\begin{figure}
\includegraphics[width=0.6\columnwidth]{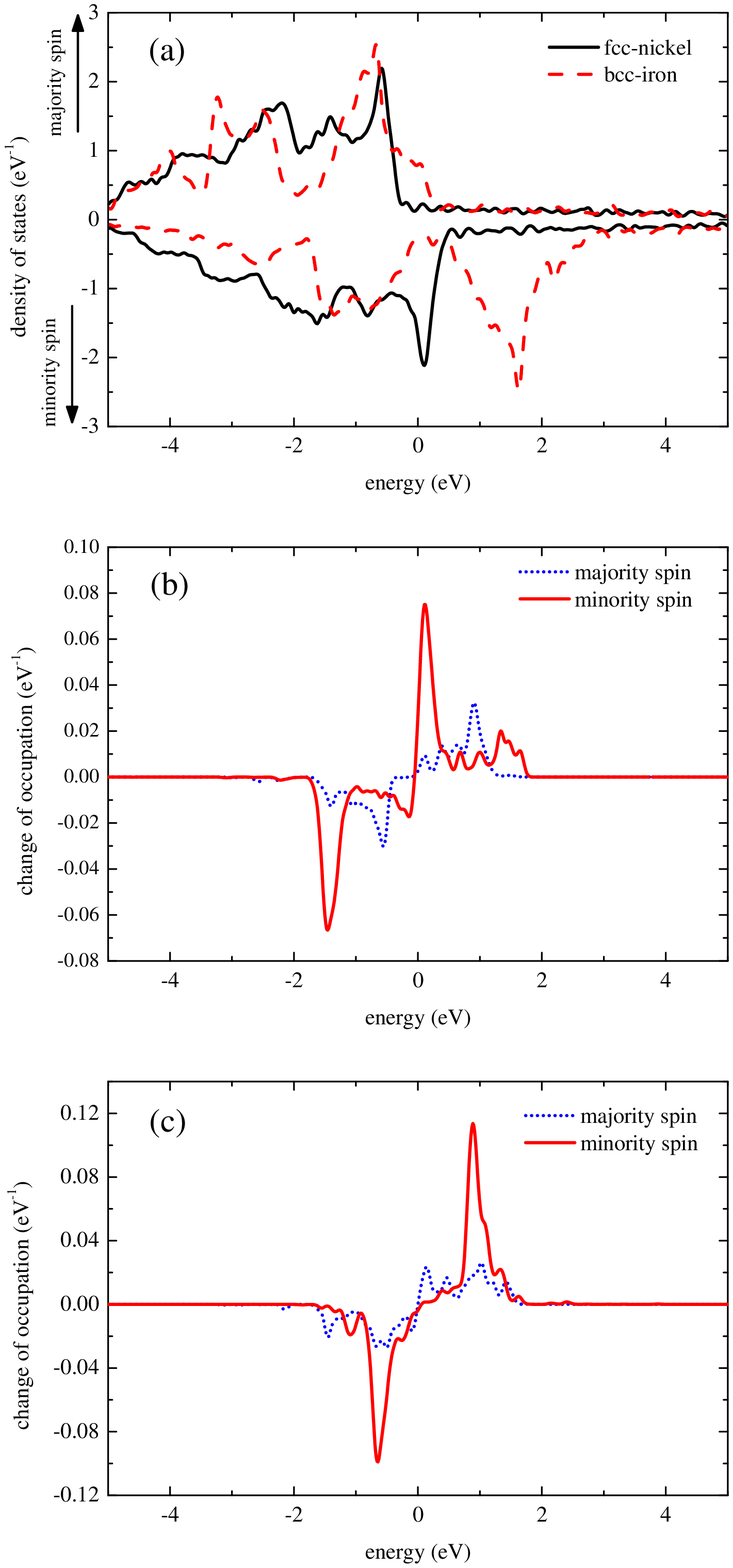}
\caption{\label{fig:changeoccopt}(Color online) Spin-resolved density of states for nickel and iron (a) as well as energy- and
  spin-resolved occupation change after the optical excitation for nickel (b) and iron (c).}
\end{figure}

As a contribution to the magnetization change, the optical excitation is negligible, but it is still interesting to take a closer look at the carrier distributions created by the laser pulse because these are
essentially the starting point of the momentum-resolved electron-phonon scattering
dynamics.  To this end, we analyze the energy-dependent occupation changes
\begin{equation}
  \Delta N_{\sigma}(\epsilon, t) = \sum_{\vec{k},\mu}
  \delta(\epsilon-\epsilon_{\vec{k}}^{\mu}) \langle\mathcal{P}_{\sigma}\rangle_{\vec{k}}^{\mu}
\left[n_{\vec{k}}^{\mu}(t)-f(\epsilon_{\vec{k}}^{\mu},T_{0})\right]
\label{DeltaN-sigma}
\end{equation}
between the dynamical distributions $n_{\vec{k}}^{\mu}(t)$ compared to the equilibrium Fermi-Dirac functions $f(\epsilon_{\vec{k}}^{\mu},T_{0})$ that describe the carrier distributions in equilibrium at the sample temperature~$T_0$ before the optical excitation. In Eq.~\eqref{DeltaN-sigma} we separate occupation changes for minority and majority spins, i.~e., for $\sigma = +$ and $-$, respectively, by projecting on the majority and minority spin contributions of the ASW wave functions using the spin-dependent weights (projections)
\begin{equation}
\langle\mathcal{P}_{\sigma}\rangle_{\vec{k}}^{\mu}=\sum_{L}
    \langle\psi_{\vec{k}}^{\mu}|\mathcal{P}_{\sigma}\psi_{\vec{k}}^{\mu}\rangle_{L}
\end{equation}
of each state $\psi_{\vec{k}}^{\mu}$ where
\begin{equation}
\begin{split}
\langle \psi_{\vec{k}}^{\mu}|\mathcal{P}_{\sigma}\psi_{\vec{k}}^{\nu}\rangle_{L}
 & =  A_{L\sigma}^{\mu*}(\vec{k})A_{L\sigma}^{\nu}(\vec{k})\langle\tilde{j}|\tilde{j}\rangle_{l\sigma}
+C_{L\sigma}^{\mu*}(\vec{k})A_{L\sigma}^{\nu}(\vec{k})\langle\tilde{h}|\tilde{j}\rangle_{l\sigma}\\
&+A_{L\sigma}^{\mu*}(\vec{k})C_{L\sigma}^{\nu}(\vec{k})\langle\tilde{j}|\tilde{h}\rangle_{l\sigma}
+C_{L\sigma}^{\mu*}(\vec{k})C_{L\sigma}^{\nu}(\vec{k})\langle\tilde{h}|\tilde{h}\rangle_{l\sigma} \ .
\end{split}
\end{equation}
The overlaps are given by
\begin{equation}
\langle\tilde{f}|\tilde{g}\rangle_{l\sigma} =\int_{0}^{r_{K}}r^{2}\tilde{f}_{l\sigma}(r)\tilde{g}_{l\sigma}(r)dr.
\label{eq:overlaps}
\end{equation}

In this paper, we always use $T_0=300$\,K as a starting point for the dynamical calculations to facilitate comparison with typical room-temperature measurements. Moreover, room temperature is still less than half the Curie temperature so that we can expect the exchange splitting to be not too different from its $T=0$\,K value, and therefore the DFT band structure should be a reasonable approximation.

Figures~\ref{fig:changeoccopt}(b) and \ref{fig:changeoccopt}(c) show the spin- and energy-resolved occupation change, computed according to Eq.~\eqref{DeltaN-sigma}, due to optical excitation at times well after the pump pulse. It corresponds to the magnetization shown in Fig.~\ref{fig:simonlyopt} at $200$\,fs. In both materials, mainly minority carriers are excited. The pronounced negative and positive spikes in the minority-spin occupation changes are separated by the photon energy 1.55\,eV and roughly coincide with maxima of the density of states [see Fig.~\ref{fig:changeoccopt}(a)] for the minority carriers. These maxima stem from the d-bands in these materials, which leads us to conclude that they play a major role in the optical excitation process. 

The information about the distribution after the optical excitation contained in Figs.~\ref{fig:changeoccopt}(b) and \ref{fig:changeoccopt}(c) allows one to draw conclusions about the maximal magnetization change achievable by
electron-phonon scattering in our model, as this distribution is the starting point for the scattering dynamics.
Electron-phonon scattering is a quasi-elastic process involving a single-electron,
i.~e., there is only a small amount of energy transferred in each
scattering event. Due to the bath assumption for the phonon system in our calculation, there are also no ``secondary
electrons'' excited because such a transfer of energy to other
electrons could only happen mediated by a phonon. We
therefore expect that electron-phonon scattering will lead to a
continuous relaxation of the excited carriers where the number of non-equilibrium electrons and holes decreases as they are scattered towards the Fermi energy. The demagnetization itself is caused by ``spin-flip'' scattering processes, i. e., scattering processes with different spin expectation values for initial and final wave functions, that occur during the relaxation process. Therefore the maximal
demagnetization that can be caused by scattering in a fixed band structure occurs when all excited majority electrons and
minority holes flip their spin while the minority electrons and
majority holes do not undergo spin-flip scattering. The relative
magnetization change in this physically rather unreasonable case is then given by
\begin{equation}
\max\frac{M}{M_{\text{eq}}}=\frac{\mu_{\text{eq}}-
2\left( N_{\text{-}}^{\text{e}}+N_{\text{+}}^{\text{h}}\right) \mu_{B}}{\mu_{\text{eq}}} \ .
\end{equation}
Here, $\mu_{B}$ is the Bohr magneton, and $M_{\text{eq}}$ and $\mu_{\text{eq}}$ denote the equilibrium
values of the material magnetization and of the magnetic moment per
unit cell, respectively. The number
of majority electrons $N_{\text{-}}^{\text{e}}$ (minority holes
$N_{\text{+}}^{\text{h}}$) per unit cell can be obtained from integrating
the occupation changes in Fig.~\ref{fig:changeoccopt} above (below)
the Fermi energy. With that estimate we find a minimal relative
magnetization due to electron-phonon scattering of 0.84 in
nickel and 0.94 in iron, which is a smaller demagnetization than observed in experiments.
Without even calculating the full dynamics, we thus expect that microscopic electron-phonon scattering with a fixed band structure
is not responsible for the pronounced drop of the magnetization observed in experiments.

\begin{figure}
\includegraphics[width=0.6\columnwidth]{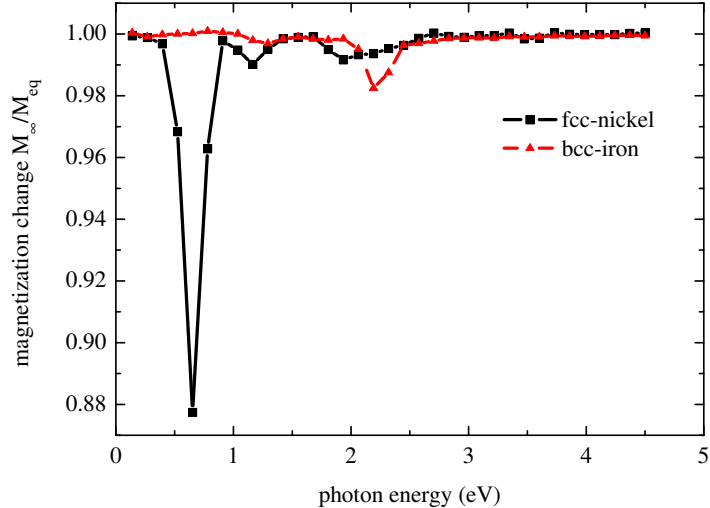}
\caption{\label{fig:optdemag_varfreq}(Color online)
  magnetization change achievable by optical excitation alone for a range of
  pump-photon energies.}
\end{figure}

We next take a closer look at the photon energy dependence of the excitation process.  Fig.~\ref{fig:optdemag_varfreq} shows the
magnetization change $M_\infty/M_\text{eq}$ vs.~the pump photon
energy for a fixed electric field amplitude of $2\times10^8$\,V/m.
For most of the pump photon energies between 0.1 and 4.5\,eV the optical excitation leads to negligible
demagnetization, as was discussed above in connection with Fig.~\ref{fig:simonlyopt} for a pump photon energy of 1.55\,eV.
Only at a pump photon energy
of about 0.7\,eV for nickel and 2.2\,eV for iron one observes a
magnetization change of significantly more than 1\,\%, because these energies correspond to the exchange splitting
of the $d$-bands in these materials, so that absorption at these
energies is likely to be associated with a change of carrier spin. The order of magnitude of our results for the achievable magnetization change by optical excitation alone seems to be in agreement with those of Zhang et al.~\cite{Zhang2009} for the change of magnetization due to the coherent excitation by an optical pulse.

\subsection{Electron-phonon scattering%
\label{sec:el-ph_scatt}} 

\begin{figure}
\includegraphics[width=0.6\columnwidth]{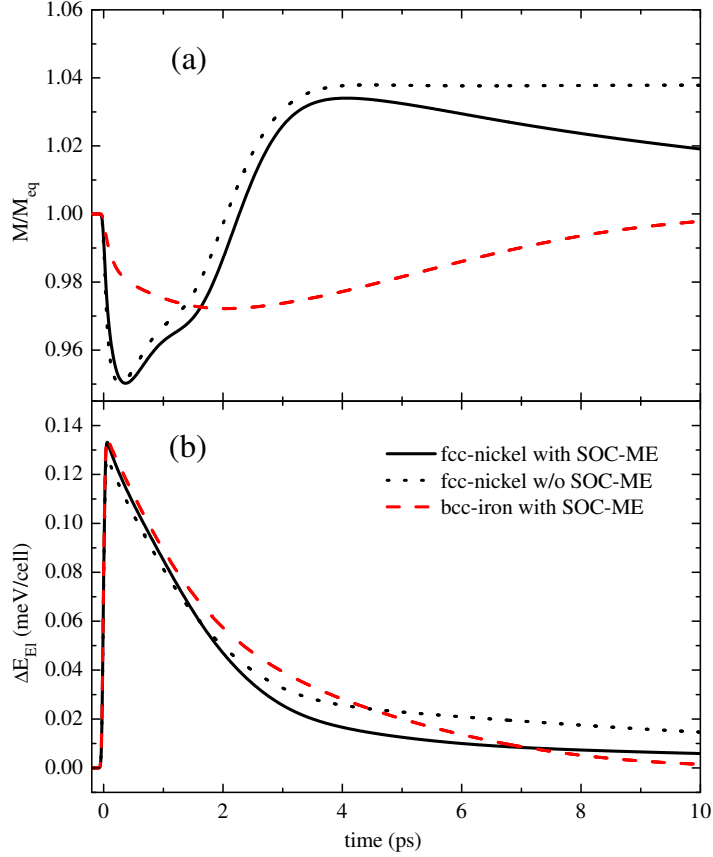}
\caption{\label{fig:scattsim155}(Color online) Normalized magnetization dynamics (a) and
   energy difference to equilibrium of the electronic
  system (b) after the optical excitation including electron-phonon
  scattering. Results obtained including the spin-orbit coupling contribution in the electron-phonon matrix element are labeled ``with SOC-ME.'' }
\end{figure}
In this section, we present results for the carrier dynamics including both optical excitation and electron-phonon scattering. We start by examining in Fig.~\ref{fig:scattsim155} the magnetization dynamics and the time evolution of the energy in the
electronic system for the same parameters that were used in Fig.~\ref{fig:simonlyopt} for the case of optical excitation only. Comparing the magnetization dynamics including electron-phonon scattering in Fig.~\ref{fig:scattsim155} to those without, cf. in Fig.~\ref{fig:simonlyopt}, one
notices a demagnetization of about 3--5\,\%. This magnetization change is smaller than the
estimate of the previous section. Due to the scattering, the dynamics now also
include a relaxation to equilibrium. This can be made visible by monitoring the energy in the electronic system
[Fig.~\ref{fig:scattsim155}(b)], which nicely shows the sudden energy
transfer from the laser pulse and a subsequent almost exponential decay
with time constants of about 2\,ps for nickel and 2.5\,ps for iron. These time constants are significantly longer than the electron-phonon coupling times obtained from the analysis of experimental data for these materials ranging from 0.3--0.5\,ps. \cite{Carpene2008,Koopmans:2005dz} Likely, this is because we neglect other scattering mechanisms (such as electron-electron scattering), which open up additional scattering paths and lead to an overall speed-up of the relaxation process. The magnetization for nickel even rises above its value at equilibrium, which is understandable because
there is no fundamental law that prohibits non-equilibrium scattering dynamics from going through intermediate states
with an increased magnetization. Whether these are reached
depends on the band structure, the properties of the
states involved, and the initial/excitation conditions. 

When comparing the calculated magnetization dynamics to experimental results, one should keep in mind that our calculation neglects changes in the band structure, i.e., the exchange splitting, and the subsequent relaxation of these changes back into equilibrium. Processes associated with the a change of the exchange splitting are expected to dominate the dynamics after a quasi-equilibrium  magnetization has been established, namely for times longer than about 5\,ps.~\cite{Djordjevic2006} A meaningful comparison with experiment of the present model should therefore be limited to a few picoseconds, which is the dynamical time scale for which the different microscopic models mentioned in the introduction have been proposed. In that time window, we find that roughly the same results (which for clarity reasons are only shown for nickel in Fig.~\ref{fig:scattsim155}) are obtained if the spin-orbit term in the interaction matrix elements [cf. Eq.~\eqref{eq:v_withso}] is neglected.

\begin{figure*}
\includegraphics[width=\textwidth]{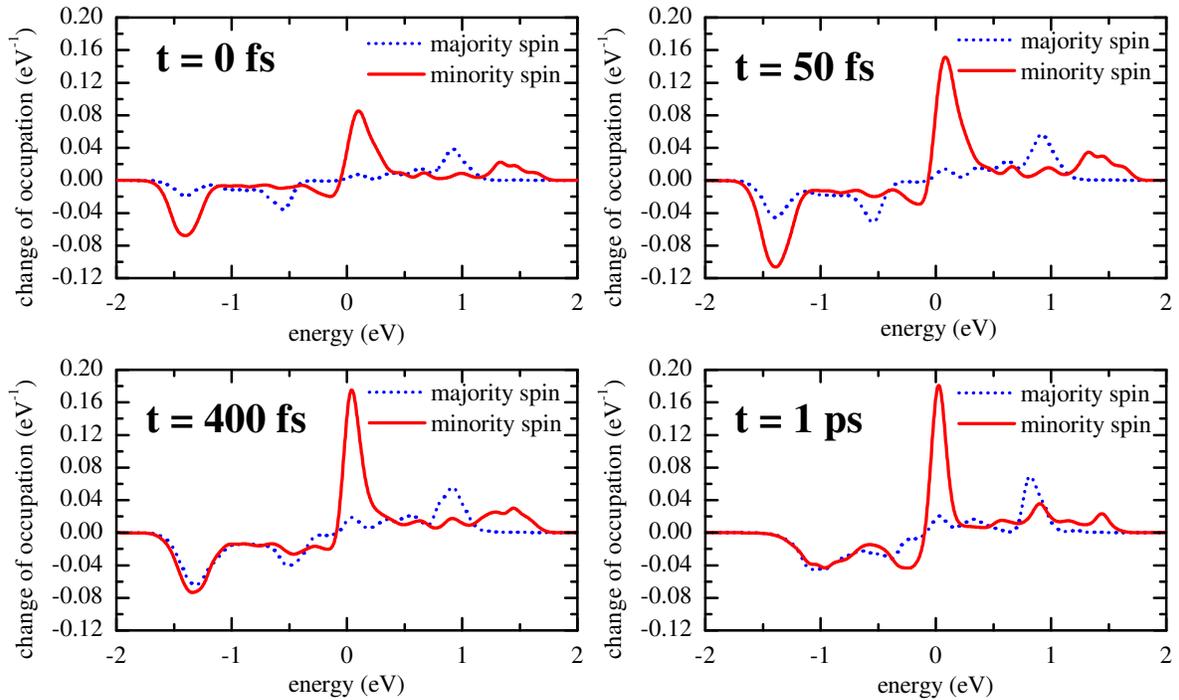}
\caption{\label{fig:diffoccen155_ni}(Color online) Energy- and
  spin-resolved occupation changes $\Delta N_{\sigma}$ at different times
   for nickel, as shown in Fig.~\ref{fig:scattsim155}, including the spin-orbit coupling in the electron-phonon matrix element. The representation is in analogy to the one for the optical excitation [Figs.~\ref{fig:changeoccopt}(b) and \ref{fig:changeoccopt}(c)].}
\end{figure*}

To get a better understanding of the demagnetization in the present model, it is
instructive to look at the carrier distributions at different stages
of the dynamics. We present only the results for nickel
in this paper, as the carrier dynamics in iron shows similar
behavior. In Fig.~\ref{fig:diffoccen155_ni} we plot the energy- and
spin-resolved occupation change at different times. Note that after the end of the optical
excitation at about 50\,fs the excited carrier density
above the Fermi energy (at 0\,eV) changes only very
slowly. In contrast, there is a strong change in the density
of holes around 1.4\,eV below the Fermi energy. It is the spin-flip of these
minority holes that leads to the demagnetization of the material in
the present model. The faster dynamics of the holes
compared to the excited electrons are due to the difference
in the density of states, cf.~Fig.~\ref{fig:changeoccopt}(a), which is considerably higher below the Fermi energy than above,
so that in this energy region there is a larger scattering phase space. In addition,  in that energy region there are ``spin hot spots,'' i.~e., points in the Brillouin zone where the states are completely spin-mixed. Their presence also contributes to spin-flip scattering processes.~\cite{Fabian1998}

\subsection{Qualitative considerations}

\begin{figure}
\includegraphics[width=0.6\columnwidth]{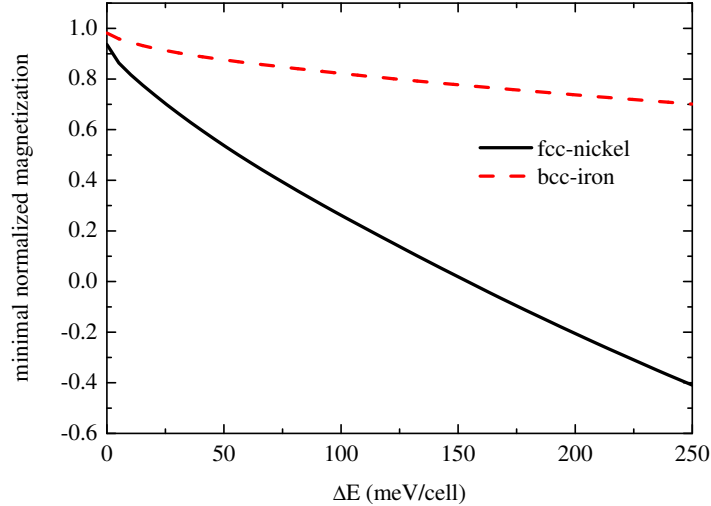}
\caption{\label{fig:maxdemag}(Color online) Theoretical
  limit for the minimal magnetization achievable by a pure redistribution in a fixed band
  structure for a range of deposited energies $\Delta E$. }
\end{figure}

As we saw from the results of the last section that electron-phonon
scattering alone cannot explain the experimentally observed
demagnetization, the next important step seems to be to extend the
existing model to other scattering mechanisms, e.g., electron-electron
or electron-impurity scattering. However, an argument based on energetics shows that their inclusion is not likely to improve the description much, if one retains the limitation that the model
contain only scattering, i.e., the redistribution of carriers in a
fixed band structure. This conclusion is based on the simple
observation that demagnetization in a fixed band structure naturally
costs energy as it requires a transfer of occupation from majority
states to minority states which are shifted up in energy by the
exchange splitting. One can make this observation quantitative by finding the
minimal magnetization that the material can attain given a fixed amount
of deposited energy $\Delta E$. This leads to a linear optimization
problem
\begin{equation}
\min_{\{ n_{\vec{k}}^{\mu}:0\leq n_{\vec{k}}^{\mu}\leq 1\}
}\sum_{\vec{k}}\!
\sum_{\mu}n_{\vec{k}}^{\mu}\left\langle S_{z}\right\rangle _{\vec{k}}^{\mu}\label{eq:optprob}
\end{equation}
with the following constraints:
\begin{subequations}
\begin{align}
\sum_{\vec{k}}\!\sum_{\mu}n_{\vec{k}}^{\mu}&=N_{\text{eq}}\\
\sum_{\vec{k}}\!\sum_{\mu}n_{\vec{k}}^{\mu}\epsilon_{\vec{k}}^{\mu}
&\leq E_{\text{eq}}+\Delta E
\end{align}
\end{subequations}
Here $N_{\text{eq}}$ denotes the total number of carriers and
$E_{\text{eq}}$ the total energy of the system in equilibrium,
i.e., before the arrival of the laser pulse. As before, the
contribution from orbital angular momentum to the total magnetization
is neglected. We solve this problem with the ab-initio results at hand for a
range of deposited energies~ $\Delta E$, and show the results
in Fig.~\ref{fig:maxdemag}. Note that we present the normalized
magnetization, i.e., the minimum obtained from the solution of
Eq.~(\ref{eq:optprob}) divided by the equilibrium magnetization because
this value can be readily compared to the 
demagnetization measured in an experiment. These values represent the minimal magnetization for
a carrier distribution in the fixed (equilibrium) band structure given
the deposited energy. It holds for all scattering mechanisms that
could be creating this distribution provided that they either conserve
energy (such as electron-electron scattering) or lead to a loss of energy
by transferring it to other systems (such as electron-phonon scattering).

By comparing the experimental demagnetization with the calculated
minimal magnetization at the amount of energy deposited in experiment
one can see whether the experimental results can, in principle, be
explained in terms of scattering alone. This comparison turns out to be not
so easy as quite a lot of parameters (e.g. the spot size, absorption,
and reflectivity) are necessary for the estimate of the deposited
energy from the measured laser intensity and some of them are
known only with a considerable uncertainty. That is why we chose to estimate the
deposited laser energy directly from the measured magnetization
dynamics. This is possible if one relies on two assumptions:
\begin{enumerate}
\item At about 5\,ps after the laser excitation the scattering
  processes have locally thermalized the material, so that the initial
  non-equilibrium dynamics that started has evolved in a  quasi-equilibrium dynamics, in which the magnetization at that time
  can be characterized by the temperature dependence of the
  magnetization in the ferromagnet $M(t\approx
  5\,\text{ps})=M(T)$ where $T=T(t\approx 5\,\text{ps})$.
\item The coupling to the substrate and other losses are so weak that
  almost all of the energy deposited by the laser is still in the
  material at that point ($t\approx 5\,\text{ps}$). However, it has
  been evenly distributed among the inner degrees of freedom.
\end{enumerate}
\begin{figure}
\includegraphics[width=0.6\columnwidth]{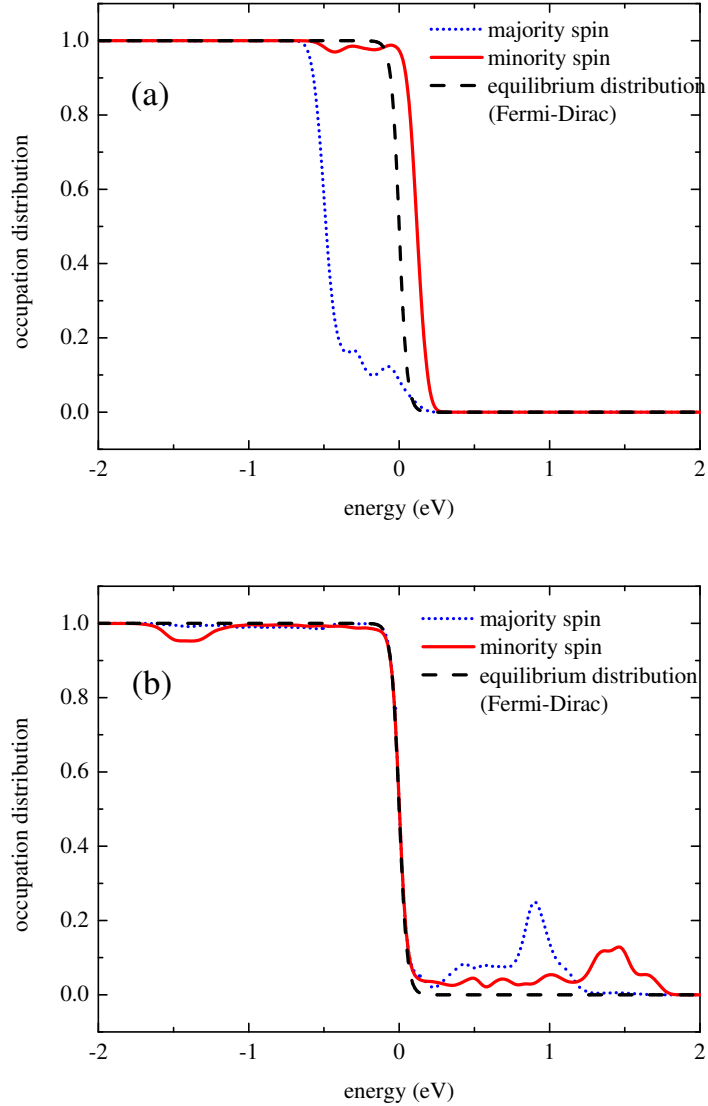}
\caption{\label{fig:distoptmaxdemag}(Color online) Energy- and
  spin-resolved occupation distributions for nickel. (a) shows the
  distributions that is necessary to attain the minimal magnetization
  for $\Delta E=100\,\text{meV}/\text{cell}$ while (b) shows a
  typical distribution that would be created by the optical excitation
  (This is a slightly different representation of the data shown in
  Fig.~\ref{fig:changeoccopt}(a). Here we show the occupation
  distribution which allows an easier comparison with the equilibrium
  distribution).}
\end{figure}
These assumptions are consistent with interpretations of measured data
by Koopmans et al.~\cite{Koopmans2000} and seem to be especially well
fulfilled for measurements on thin films. They can now be used to
extract the deposited energy from the measured magnetization at 5\,ps, and to read off the corresponding achievable minimum magnetization from Fig.~\ref{fig:distoptmaxdemag}.
This can be compared with the ``quenched'' magnetization reached in the
same measurement. In typical data for nickel \footnote{Measurements~\cite{Koopmans:2010ds} on
  Ni (15\,nm) film with 5.0\,mJ/cm$^2$} and iron \footnote{Measurements~\cite{Carpene2008}
  on Fe (7\,nm) film with 6.0\,mJ/cm$^2$} at high intensities we find
for the normalized magnetization after thermalization values of
$M_{\text{Ni}}(5\,\text{ps})=0.1$ and
$M_{\text{Fe}}(5\,\text{ps})=0.8$. Using the equilibrium temperature dependence of the magnetization $M(T)$, we conclude that the
temperature after local thermalization is about 625\,K for nickel and
about 800\,K for iron, respectively. As we assume an even
distribution among the material's degrees of freedom, we can calculate
the deposited energy as an integral over the heat capacity
$C_{\text{p}}(T)$:
\begin{equation}
\Delta E=\int_{300\,\text{K}}^{T(5\,\text{ps})}dTC_{\text{p}}(T)
\end{equation}
which we solved using experimental data for $C_{\text{p}}(T)$
\cite{Landolt-Bornstein1986} yielding $\Delta
E(\text{Ni})=100\,\text{meV}/\text{cell}$ and $\Delta
E(\text{Fe})=160\,\text{meV}/\text{cell}$. For these energies, Fig.~\ref{fig:maxdemag}
yields 0.26 and 0.77 as minimal achievable magnetizations for nickel and iron, respectively. These values should be compared
to the experimentally observed quenched magnetizations of 0.1 for nickel and 0.7 for
iron. As the experimentally measured minima only slightly violate the
theoretical bounds, one could be inclined to conclude that this argument
does not rule out a demagnetization on the basis of pure redistribution
in a fixed band structure. That view changes, however, if one looks at
the corresponding distribution functions that are necessary to attain
the theoretical magnetization minima. The one for nickel is shown in
Fig.~\ref{fig:distoptmaxdemag}(a) and should be compared to the
distribution that is created by pure optical excitation
[Fig.~\ref{fig:distoptmaxdemag}(b)]. As discussed before, the
optically excited distribution is the starting point for all
scattering processes and we would expect these to bring the system
back to a Fermi-Dirac distribution (at a higher temperature) which is
also displayed in the figure. It is not at all likely that in the
course of this process there will be an intermediate state that has a
distribution that is anywhere close to the one shown in
Fig.~\ref{fig:distoptmaxdemag}(a) for two reasons: First, for a
magnetization close to the theoretical limit a highly ``ordered''
distribution is necessary, which is unlikely to be reached by random
scattering processes. Second, the state shown in
Fig.~\ref{fig:distoptmaxdemag}(a) lies very far off from the direct
continuous transition from the distribution in
Fig.~\ref{fig:distoptmaxdemag}(b) to the equilibrium distribution,
both in terms of a simple relaxation time approximation
and if one considers a quasi-elastic process, such as electron-phonon
scattering, where we have a slow, but continuous energy relaxation of
the exited carriers towards the Fermi energy where eventually
non-equilibrium electrons and holes cancel out.

Even though this argument is not a rigorous, we find it convincing enough to draw the conclusion that scattering dynamics in a fixed band
structure cannot explain the observed ultrafast demagnetization. It is therefore important to include the dynamical changes in the band structure, i.e., the exchange splitting, in a comprehensive microscopic theory of ultrafast demagnetization in ferromagnets.

\section{Conclusions}

The main objective of this paper was to analyze in detail the dynamics due to one of the proposed mechanisms for ultrafast demagnetization: the Elliott-Yafet process based on electron-phonon scattering. To this end, we carried out a numerical analysis without adjustable parameters including the laser excitation and the scattering dynamics on the level of Boltzmann scattering integrals. We evaluated the model for the elementary ferromagnets nickel and iron utilizing realistic band structures and matrix elements obtained from ab-initio calculations. As in previous studies,~\cite{Krauss:2009gc,Steiauf:2009ca} we kept the band structure fixed. In this case, the computed demagnetization for realistic pump-laser intensities is smaller by almost a factor of ten than what is observed in experiments. An additional argument shows that this bound for the achievable magnetization ``quenching'' is likely to hold as well for other scattering mechanisms, such as electron-electron or electron-impurity scattering. We interpret our numerical results that any fully microscopic model that tries to explain ultrafast demagnetization by scattering dynamics really should include a dynamical change of the band structure, i.e., the exchange splitting. A microscopic determination of this change seems more important for the understanding of the demagnetization process than studies focussing on Elliott-Yafet-type mechanisms.

\acknowledgments 

We are grateful to J.~K\"ubler for providing us with
his ASW density-functional code and introducing us to details of the
implementation. We benefitted from discussions with him and
C.~Ambrosch-Draxl regarding density-functional theory in general,
as well as with P.~Oppeneer and K.~Carva regarding ab-initio calculations for
ferromagnets.

\appendix*

\section{Numerical Method}
\label{sec:nummeth}
For the numerical solution of Eq.~(\ref{eq:dndt}) we replace the
energy delta function in the scattering rates, Eq.~(\ref{eq:def_w}),
by a gaussian of finite width to allow a Brillouin zone
integration on the chosen grid of $\vec{k}$-points. The FWHM of this
broadened distribution is taken to be 15\,meV for the DFT grid at hand, 
and convergence of the results with respect to grid size and
distribution width was checked.

\subsection{Reducing the dimensionality of the problem}
Two simplifications help to reduce the numerical effort for the
solution of Eq.~(\ref{eq:dndt}):

The first is based on the fact that only the states in a limited
energy range around the Fermi energy will experience an occupation
change in the course of the dynamics. Due to the structure of the
equilibrium distribution
\begin{equation}
(n_{\vec{k}}^{\mu})_{\text{eq}}=\frac{1}{e^{\left(\epsilon_{\vec{k}}^{\mu}-\mu\right)/k_{B}T_{0}}+1}
\end{equation}
the states far ($\geq 200\,\text{meV}$ for $T_0=300\,\text{K}$) above
the Fermi energy ($\mu\approx E_F$) will be empty while those far
below the Fermi energy will be fully occupied.  As the incoming laser
pulse will only cause resonant transitions between occupied and unoccupied
states, the occupation change due to
the optical excitation is limited to an energy range around the Fermi
energy. This is clearly seen in the energy resolved occupation change
due to the optical excitation in Fig.~\ref{fig:changeoccopt}. States
far away from the Fermi energy
($\bigl|\epsilon_{\vec{k}}^{\mu}-E_F\bigr|\gg E_{\text{las}}$) remain
at their equilibrium values. As the electron-phonon scattering
transfers only small amounts of energy in each scattering process, the
occupation of these states is not influenced by the following
scattering dynamics either. That is why one can safely assume the
occupation of these states to remain constant in time. Only states in
an energy range $\bigl|\epsilon_{\vec{k}}^{\mu}-E_F\bigr|\leq\delta E$
are actually included in the dynamical calculation of the occupation
numbers. So, for each $\vec{k}$-point, we only include the subset of
bands $n_{\vec{k}}^{\mu_j}$ into the dynamical calculation that fall
in the chosen energy range $\delta E$. For the calculations including
only the optical excitation in Sec.~\ref{sec:opt_exc} we took
$\delta E$ to be 5\,eV, but for the photon energy under
investigation (1.55\,eV) a much smaller range actually suffices. We
therefore reduced it to 2\,eV for the calculations including scattering in
Sec.~\ref{sec:el-ph_scatt}.

The second simplification can be made due to the crystal symmetries of
the materials under investigation. These symmetries imply that the
wave functions of two states at different $\vec{k}$-points which are
related by a crystal symmetry operation are also connected by the same
symmetry operation. From that one can deduce that the modulus of an
electron-phonon matrix element between two states does not change if
one applies a crystal symmetry operation on the initial and the final
state. In other words, electron-phonon scattering will not break the
symmetry of an occupation distribution that has the same symmetry as
the crystal (e.g., the thermal distribution before optical excitation). The optical
excitation could break that symmetry as it involves the
scalar product with an external electric field. In our paper, we
restrict ourselves to the description of the typical experimental case
where the laser field is parallel to the material magnetization so
that the symmetry of the occupation distribution is not broken by the
optical excitation. In this case, it is not necessary to compute all
occupation numbers $n_{\vec{k}}^{\mu}$. Rather all information about the
occupation distribution is contained in any subset $\{\vec{k}_j\}$ of
$\vec{k}$-points which constitutes an irreducible wedge of the Brillouin zone.

\subsection{Numerical evaluation}

With these two simplifications we are left with a subset of occupation
numbers $n_j$ which need to be calculated dynamically. Here
$j=(\vec{k}_j,\mu_j)$ denotes a multi-index that includes the
$\vec{k}$-point as well as the band index of the state. With the help
of this notation, Eq.~(\ref{eq:dndt}) can be reformulated to yield
\begin{equation}
\frac{\partial n_{j}}{\partial t}=\left.\frac{\partial n_{j}}{\partial t}\right|_{\text{opt}}+\left.\frac{\partial n_{j}}{\partial t}\right|_{\text{e-p}}
=\left|\vec{E}(t)\right|^{2}\sum_{i}\mathbf{B}_{ji}^{\text{opt}}n_{i}
+\sum_{i}\left(n_{j}\mathbf{A}_{ji}^{\text{e-p}}\left(1-n_{i}\right)-\left(1-n_{j}\right)\mathbf{A}_{ij}^{\text{e-p}}n_{i}\right),
\end{equation}
where all $n_j$-independent quantities are contained in the
constant matrices $\mathbf{A}^\text{e-p}$ and $\mathbf{B}^\text{opt}$,
which can be precomputed for the chosen set of states. Interpreting
the occupation numbers $n_j$ as a vector $\vec{n}$ we can
alternatively write this as
\begin{equation}
\frac{\partial}{\partial t}\vec{n}=\left|\vec{E}(t)\right|^{2}\mathbf{B}^{\text{opt}}\vec{n}
+\text{diag}\left(\vec{n}\right)\mathbf{A}^{\text{e-p}}\left(\vec{1}-\vec{n}\right) 
-\text{diag}\left(\vec{1}-\vec{n}\right)\left(\mathbf{A}^{\text{e-p}}\right)^{T}\vec{n} 
\label{eq:dndt_mat}
\end{equation}
where $\text{diag}(\vec{n})$ is a matrix with the occupation numbers
on the diagonal. $\vec{1}$ denotes a vector with all entries set to
one. In the form of Eq.~(\ref{eq:dndt_mat}) the differential equation
is especially well suited for a numerical evaluation. We used a MATLAB
algorithm~\cite{Shampine1997} for the solution.

\bibliography{e-pn_demag}
\end{document}